\documentclass[letterpaper]{ptephy_v1}

\usepackage{boites}

\usepackage{xspace}
\usepackage{graphicx}
\usepackage{lineno}
\usepackage{setspace}

\newcommand{\kpnnEq}{K_{L} \to \pi^0 \nu \overline{\nu}}
\newcommand{\kpnn}{$\kpnnEq$\xspace}
\newcommand{\br}{Br$(\kpnnEq)<$\xspace}
\newcommand{\kpluspnn}{$K^{+} \to \pi^{+} \nu \overline{\nu}$\xspace}
\newcommand{\p}{$\pi^{0}$\xspace}
\newcommand{\ps}{$\pi^{0}$'s\xspace}

\newcommand{\KL}{$K_{L}$\xspace}
\newcommand{\KLs}{$K_{L}$'s\xspace}
\newcommand{\kppp}{$K_{L} \to 3 \pi^{0}$\xspace}
\newcommand{\kcppp}{$K_{L} \to \pi^{+} \pi^{-} \pi^{0}$\xspace}
\newcommand{\kppEq}{K_{L} \to 2 \pi^{0}}
\newcommand{\kpp}{$\kppEq$\xspace}
\newcommand{\kpxEq}{K_{L} \to \pi^{0} X^{0}}
\newcommand{\kpx}{$\kpxEq$\xspace}
\newcommand{\ke}{$K_{L} \to \pi^{\pm} e^{\mp} \nu_{e} $\xspace}
\newcommand{\kmu}{$K_{L} \to \pi^{\pm} \mu^{\mp} \nu_{\mu} $\xspace}
\newcommand{\kgg}{$K_{L} \to 2 \gamma$\xspace}

\newcommand{\pt}{$P_{\mathrm{T}}$\xspace}
\newcommand{\zvtx}{$Z_{\mathrm{vtx}}$\xspace}

\newcommand{\myPRL}[1]{Phys. Rev. Lett. \textbf{#1}\xspace}
\newcommand{\myPRD}[1]{Phys. Rev. D \textbf{#1}\xspace}

\newcommand{\myPLB}[1]{Phys. Lett. B \textbf{#1}\xspace}
\newcommand{\myPTEP}[1]{Prog. Theor. Exp. Phys. \textbf{#1}\xspace}

\newcommand{\myJHEP}[1]{J. High Energy Phys. \textbf{#1}\xspace}
\newcommand{\myNIMA}[1]{Nucl. Instrum. Methods A \textbf{#1}\xspace}
\newcommand{\myetal}{\textit{et al.}\xspace}

\begin{document}


\title{A new search for the \kpnn and \kpx decays}

\renewcommand{\thefootnote}{\fnsymbol{footnote}}
\author{\collaborator{J-PARC KOTO collaboration}
\name{J.~K.~Ahn}{1},
\name{K.~Y.~Baek}{2},
\name{S.~Banno}{3},
\name{B.~Beckford}{4},
\name{B.~Brubaker}{5,18},
\name{T.~Cai}{5,19},
\name{M.~Campbell}{4},
\name{C.~Carruth}{4,20},
\name{S.~H.~Chen}{6},
\name{S.~Chu}{5},
\name{J.~Comfort}{7},
\name{Y.~T.~Duh}{6},
\name{T.~Furukawa}{8},
\name{H.~Haraguchi}{3},
\name{T.~Hineno}{9},
\name{Y.~B.~Hsiung}{6},
\name{M.~Hutcheson}{4},
\name{T.~Inagaki}{10},
\name{M.~Isoe}{3},
\name{E.~Iwai}{3,21},
\name{T.~Kamibayashi}{11},
\name{I.~Kamiji}{9},
\name{N.~Kawasaki}{9},
\name{E.~J.~Kim}{12},
\name{Y.~J.~Kim}{13},
\name{J.~W.~Ko}{13},
\name{T.~K.~Komatsubara}{10,22},
\name{A.~S.~Kurilin}{14,23},
\name{G.~H.~Lee}{12},
\name{H.~S.~Lee}{15}, 
\name{J.~W.~Lee}{3,24}
\name{S.~K.~Lee}{12},
\name{G.~Y.~Lim}{10,22},
\name{C.~Lin}{6},
\name{J.~Ma}{5},
\name{Y.~Maeda}{9\ast,25},
\name{T.~Masuda}{9,26},
\name{T.~Matsumura}{16},
\name{D.~Mcfarland}{7},
\name{J.~Micallef}{4,27},
\name{K.~Miyazaki}{3},
\name{K.~Morgan}{5,28},
\name{R.~Murayama}{3},
\name{D.~Naito}{9,29},
\name{K.~Nakagiri}{9},
\name{Y.~Nakajima}{9,30},
\name{Y.~Nakaya}{3,23},
\name{H.~Nanjo}{9,31},
\name{T.~Nomura}{10,22},
\name{T.~Nomura}{11},
\name{Y.~Odani}{8},
\name{R.~Ogata}{8},
\name{H.~Okuno}{10},
\name{T.~Ota}{8},
\name{Y.~D.~Ri}{3},
\name{M.~Sasaki}{11},
\name{N.~Sasao}{17},
\name{K.~Sato}{3,30},
\name{T.~Sato}{10},
\name{S.~Seki}{9},
\name{T.~Shimogawa}{8,29},
\name{T.~Shinkawa}{16},
\name{S.~Shinohara}{9},
\name{K.~Shiomi}{3,32},
\name{J.~S.~Son}{12},
\name{J.~Stevens}{7,33},
\name{S.~Su}{4},
\name{Y.~Sugiyama}{3,29},
\name{S.~Suzuki}{8},
\name{Y.~Tajima}{11},
\name{G.~Takahashi}{9},
\name{Y.~Takashima}{3},
\name{M.~Tecchio}{4},
\name{I.~Teo}{5,34},
\name{M.~Togawa}{3},
\name{T.~Toyoda}{3},
\name{Y.~C.~Tung}{6,35},
\name{T.~Usuki}{9},
\name{Y.~W.~Wah}{5},
\name{H.~Watanabe}{10,22},
\name{N.~Whallon}{4,36},
\name{J.~K.~Woo}{13},
\name{J.~Xu}{4},
\name{M.~Yamaga}{3,37},
\name{S.~Yamamoto}{8},
\name{T.~Yamanaka}{3},
\name{H.~Yamauchi}{8},
\name{Y.~Yanagida}{3},
\name{H.~Yokota}{16},
\name{H.~Y.~Yoshida}{11},
and \name{H.~Yoshimoto}{3}}

\address{
\affil{1}{$^{1}$Department of Physics, Korea University, Seoul 02841, Republic of Korea}\\
\affil{2}{$^{2}$Department of Physics, Pusan National University, Busan 46241, Republic of Korea}\\
\affil{3}{$^{3}$Department of Physics, Osaka University, Toyonaka, Osaka 560-0043, Japan}\\
\affil{4}{$^{4}$Department of Physics, University of Michigan, Ann Arbor, MI 48109, USA}\\
\affil{5}{$^{5}$Enrico Fermi Institute, University of Chicago, Chicago, IL 60637, USA}\\
\affil{6}{$^{6}$Department of Physics, National Taiwan University, Taipei, Taiwan 10617, Republic of China}\\
\affil{7}{$^{7}$Department of Physics, Arizona State University, Tempe, AZ 85287, USA}\\
\affil{8}{$^{8}$Department of Physics, Saga University, Saga 840-8502, Japan}\\
\affil{9}{$^{9}$Department of Physics, Kyoto University, Kyoto 606-8502, Japan}\\
\affil{10}{$^{10}$Institute of Particle and Nuclear Studies, High Energy Accelerator Research Organization (KEK), Tsukuba, Ibaraki 305-0801, Japan}\\
\affil{11}{$^{11}$Department of Physics, Yamagata University, Yamagata 990-8560, Japan}\\
\affil{12}{$^{12}$Division of Science Education, Chonbuk National University, Jeonju 54896, Republic of Korea}\\
\affil{13}{$^{13}$Department of Physics, Jeju National University, Jeju 63243, Republic of Korea}\\
\affil{14}{$^{14}$Laboratory of Nuclear Problems, Joint Institute for Nuclear Researches, Dubna, Moscow reg. 141980, Russia}\\
\affil{15}{$^{15}$RISP, Institute for Basic Science, Daejeon 34047, Republic of Korea}\\
\affil{16}{$^{16}$Department of Applied Physics, National Defense Academy, Kanagawa 239-8686, Japan}\\
\affil{17}{$^{17}$Research Institute for Interdisciplinary Science, Okayama University, Okayama 700-8530, Japan}\\
\affil{18}{$^{18}$Present address: Department of Physics, Yale University, New Haven, CT 06520-8120, USA}\\
\affil{19}{$^{19}$Present address: Department of Physics and Astronomy, University of Rochester, Rochester, NY 14627-0171, USA}\\
\affil{20}{$^{20}$Present address: Department of Physics, University of California at Berkeley, Berkeley, CA 94720-7300, USA}\\
\affil{21}{$^{21}$Present address: Department of Physics, University of Michigan, Ann Arbor, MI 48109, USA}\\
\affil{22}{$^{22}$Also at J-PARC Center, Tokai, Ibaraki 319-1195, Japan}\\
\affil{23}{$^{23}$Deceased}\\
\affil{24}{$^{24}$Present address: Department of Physics, Korea University, Seoul 02841, Republic of Korea}\\
\affil{25}{$^{25}$Present address: Kobayashi-Maskawa Institute, Nagoya University, Nagoya 464-8602, Japan}\\
\affil{26}{$^{26}$Present address: Research Institute for Interdisciplinary Science, Okayama University, Okayama 700-8530, Japan}\\
\affil{27}{$^{27}$Present address: Department of Physics and Astronomy, Michigan State University, East Lansing, MI 48824, USA}\\
\affil{28}{$^{28}$Present address: Department of Physics, University of Wisconsin-Madison, Madison, WI 53706-1390, USA}\\
\affil{29}{$^{29}$Present address: Accelerator Laboratory, High Energy Accelerator Research Organization (KEK), Tsukuba, Ibaraki 305-0801, Japan}\\
\affil{30}{$^{30}$Present address: Kamioka Observatory, Institute for Cosmic Ray Research, University of Tokyo, Kamioka, Gifu 506-1205, Japan}\\ 
\affil{31}{$^{31}$Present address: Department of Physics, Osaka University, Toyonaka, Osaka 560-0043, Japan}\\
\affil{32}{$^{32}$Present address: Institute of Particle and Nuclear Studies, High Energy Accelerator Research Organization (KEK), Tsukuba, Ibaraki 305-0801, Japan}\\
\affil{33}{$^{33}$Present address: Department of Physics, Cornell University, Ithaca, NY 14853-2501, USA}\\
\affil{34}{$^{34}$Present address: Department of Physics, University of Illinois at Urbana-Champaign, Urbana, IL 61801-3080, USA}\\
\affil{35}{$^{35}$Present address: Enrico Fermi Institute, University of Chicago, Chicago, IL 60637, USA}\\
\affil{36}{$^{36}$Present address: Department of Physics, University of Washington, Seattle, WA 98195-1560, USA}\\
\affil{37}{$^{37}$Present address: Controls and Computing Division, Japan Synchrotron Radiation Research Institute (JASRI), Hyogo 679-5198, Japan}\\
\email{maeda\_y@scphys.kyoto-u.ac.jp}
}

\begin{abstract}%
We searched for the $CP$-violating rare decay of neutral kaon, \kpnn,
in data from the first 100 hours of physics running in 2013 of the J-PARC KOTO experiment. 
One candidate event was observed
while $0.34\pm0.16$ background events were expected.
We set an upper limit of $5.1\times10^{-8}$ for the branching fraction at the 90\% confidence level (C.L.).
An upper limit of $3.7\times10^{-8}$ at the 90\% C.L. for the \kpx decay was also set for the first time,
where $X^{0}$ is an invisible particle with a mass of 135 MeV/$c^{2}$.
\end{abstract}%

\subjectindex{C03, C30}

\maketitle

\renewcommand{\thefootnote}{\arabic{footnote}}

\section{Introduction}
The rare decay \kpnn of the long-lived neutral kaon is a direct $CP$-violating process \cite{kpnn_Littenberg,KL_review},
and is one of the most sensitive probes to search for new physics beyond the standard model (SM) of particle physics.
Because this decay proceeds through the flavor changing neutral current via the $s \rightarrow d$ transition,
it is strongly suppressed in the SM and is sensitive to new heavy particles contributing to the decay \cite{SUSY,Zprime}.
The SM predicts the branching fraction (Br) to be $(3.00\pm0.30)\times10^{-11}$ \cite{SMBr},
while the current experimental upper limit is $2.6\times10^{-8}$ at the 90\% confidence level (C.L.)
set by the KEK E391a experiment \cite{E391a}.
The BNL E949 experiment \cite{E949} has set an indirect and model-independent limit \br$1.46\times10^{-9}$
based on the measurement of the \kpluspnn branching fraction \cite{GNLimit}.

The signature of the \kpnn decay is a single \p from a \KL decay in flight without any other detectable particles.
The experimental study is also sensitive to the two-body decay \kpx, where $X^{0}$ is an invisible boson. 
It was recently pointed out that the limit $1.46\times10^{-9}$ based on the $K^{+}$ measurement
does not apply to the \kpx decay if the mass of $X^{0}$ is close to the \p mass
and new physics with a weakly-interacting light particle can be probed through the \kpnn study \cite{KLpi0X, KLpi0X_new}.

The KOTO experiment \cite{KOTO} at the Japan Proton Accelerator Research Complex (J-PARC) \cite{J-PARC}
is the successor of E391a, which was the first dedicated search for \kpnn.
KOTO adopts the same experimental techniques
and aims at a sensitivity of $10^{-11}$ by adding a new beam line and various improvements of the detector.
This letter reports the results from the first physics data collected in May 2013.

\section{Apparatus}
The experiment was conducted at the Hadron Experimental Facility (HEF) \cite{HEF} of J-PARC.
A beam of 30-GeV protons was slowly extracted from the Main Ring (MR) accelerator \cite{MR}
onto a 66-mm-long gold target \cite{AuTarget} at HEF. 
The \KLs produced at an angle of 16$^{\circ}$ direction from the proton beam were transported through a neutral beam line \cite{beamline, KLyield}
consisting of two collimators made of iron and tungsten, a sweeping magnet, and a 7-cm-thick lead photon absorber.
The solid angle of the neutral beam after collimation was 7.8~$\mu$sr,
and its size was $8 \times 8$~cm$^{2}$ at 20~m downstream from the target.
The peak \KL momentum was 1.4~GeV/$c$.
The beam also contained neutrons and photons.
Neutrons outside the nominal beam solid angle, which came from scattering inside the collimators, are referred to as ``halo neutrons.''

\begin{figure}[tb]
	\begin{center}
		\includegraphics[width=0.95\textwidth,bb=0 0 712 208]{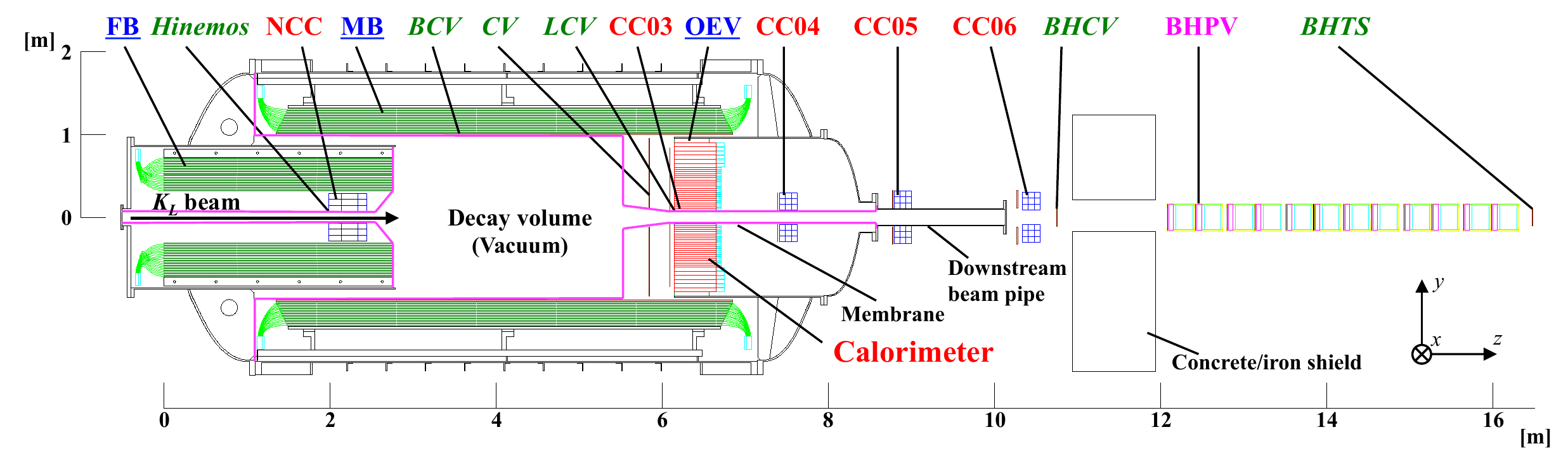}
         \end{center}
	\caption{Cut-out-view of the KOTO detector assembly in the physics run of May 2013.
		Veto counters with their names written
		in blue underlined, green italic and red regular letters
		are made of lead-scintillator sandwich, plastic scintillator and undoped CsI crystal, respectively.
		The downstream counter named as BHPV is made of lead-aerogel sandwich.
		The Hinemos and BHTS counters were used only for background studies and calibration, but not for veto.}
	\label{fig:detector}
\end{figure}

A schematic view of the detector is shown in Fig.~\ref{fig:detector}.
The 3-m-long decay volume along the beam direction was inside the detector.
Two photons from a \p decay were detected by the electromagnetic calorimeter.
It consisted of 2716 undoped CsI crystals, stacked inside a 1.9-m-diameter cylinder except the central 20$\times$20~cm$^{2}$ region. 
The size of each crystal within (outside) the central 1.2$\times$1.2~m$^{2}$ region
was 2.5$\times$2.5~cm$^{2}$ (5.0$\times$5.0~cm$^{2}$) in cross section and 50~cm in length,
which corresponded to 27 radiation lengths in the beam direction.
The energy resolution of the calorimeter was evaluated as $\sigma_{E}/E = (0.99\oplus1.74/\sqrt{E})$\%,
where $\oplus$ indicates a quadratic sum, and $E$ is in GeV \cite{SatoPhD}.
Details of the calorimeter are described in Refs.~\cite{CW,IwaiNIM}.
Subsystems other than the calorimeter in the KOTO detector were ``veto counters,''
which ensured that no other detectable particles were emitted in the \KL decay.
The gaps between the cylinder and the crystals of the calorimeter were filled with lead-scintillator sandwich counters named OEV \cite{OEV}.
The beam hole of $15\times15$~cm$^{2}$ was located at the center of the calorimeter to let the beam particles pass through.
The hole was surrounded with counters named LCV and CC03,
which were composed of plastic scintillator plates and CsI crystals, respectively.
The upstream surface of the calorimeter was covered with the Charged Veto (CV) counter \cite{CV},
consisting of two layers of 3-mm-thick plastic scintillator sheets, to veto the \KL decays with charged particles.
The outside region of the decay volume was surrounded by the Main Barrel (MB) \cite{MB}.
The MB was a sandwich-type shower counter with lead and plastic scintillator sheets,
and detected extra particles from the \kppp and \kpp decays.
A counter of plastic scintillator sheets, named BCV, covered the inner surface of MB.
The upstream end of the decay region was covered by the Front Barrel (FB) \cite{MB} and the Neutron Collar Counter (NCC) \cite{NCC}.
The FB was a lead-scintillator sandwich counter and NCC was made of CsI crystals. 
These veto counters detected photons from \KL decays in the upstream direction.
The downstream region of the calorimeter was covered by a series of photon veto counters named CC04, CC05 and CC06,
and in-beam counters named BHCV and BHPV \cite{BHPV}.
They were meant to detect particles escaping in the forward direction through the beam hole of the calorimeter.
Counters except for CC05, CC06, BHCV, and BHPV were located in vacuum at 0.1~Pa,
and the decay volume was kept at $5\times10^{-5}$~Pa
to suppress \ps produced by the interactions of neutrons in the beam with residual gas.
This high vacuum region was separated from the counter region with a multi-layer film called ``membrane,''
whose area density was 202~$\mu$g/cm$^{2}$.  

Signals from the calorimeter and the veto counters were recorded as waveforms
to separate multiple overlapping hits.
For most detectors, 125-MHz waveform digitizers \cite{FADC125} were used
after shaping the raw pulses to Gaussian-like pulses with $\sigma\sim$30~ns.
For the in-beam counters, whose counting rates reached an order of MHz,
500-MHz digitizers \cite{FADC500} without pulse shaping were used.

\section{Data taking}
The data analyzed in this letter was collected in 100 hours between May 19 and 23, 2013,
before the data taking was suspended due to an incident at HEF caused by an accelerator malfunction \cite{accident}.
The beam power incident on the gold target was 24~kW, which corresponded to $3\times10^{13}$ protons on target (POT)
in a 2-second-long duration (spill) with a 6 second repetition.
The total number of POT was $1.6\times10^{18}$.
The data acquisition system was triggered by two levels of trigger logic \cite{DAQ}.
The first level trigger (L1) required the total energy detected in the calorimeter to be larger than 550~MeV.
The acceptance loss due to this and the subsequent offline requirement was less than 1\%.
It also required the total energy deposition to be less than 1~MeV in CV
and less than 50-60~MeV in NCC, MB, and CC03
to reject most of \KL decays with charged particles or photons hitting the veto counters.
The energy thresholds for the L1 selection were set to be looser than those for the offline analysis.
In the second level trigger (L2), the center of energy deposition (COE) in the calorimeter was calculated
and the distance from the beam center to the COE position was required to be larger than 165~mm.
This requirement was used to suppress the \kppp decay
because the energy of photons from the decay will balance in the transverse direction,
while the COE distance tends to be large in the \kpnn decay due to the neutrinos in the final state.
With these trigger requirements, 300 million events were collected for \kpnn.
To collect data from the \kppp, \kpp, and \kgg decays simultaneously for normalization and calibration purposes,
disregarding the L2 decision, events which satisfied the L1 requirements, and L1 requirements without veto
were also recorded with prescaling factors of 30 and 300, respectively.
In total, $\sim$8000 events per spill were collected.

\section{Reconstruction and selection}
At first, photon candidates in the calorimeter were reconstructed.
After requiring an energy deposit larger than 3~MeV to identify hits in a single crystal,
a cluster of hits in adjacent crystals was associated to an electromagnetic shower generated by a photon as described in Ref.~\cite{MasudaKL}.
After the photon reconstruction,
events with two photons were selected for \p reconstruction.
The \p-decay vertex position (\zvtx) was obtained assuming that a $\pi^{0}\to2\gamma$ decay occurred on the beam ($z$) axis.
Here, the $z$ axis is defined as the center of the beam as shown in Fig.~\ref{fig:detector}
and $z=0$~m corresponds to 21.5~m downstream from the target.
The transverse momentum (\pt) and the decay time of the \p were calculated with the calorimeter information and \zvtx.

A series of event selection criteria (cuts) was imposed on the reconstructed \p kinematics and the hit information of the veto counters.
To ensure the L1 and L2 requirements of the calorimeter in the offline analysis were met,
the sum of the two photon energies was required to be larger than 650~MeV,
and the COE position was required to be farther than 200~mm from the beam center.
Events were rejected if any channel in the veto counters had a hit coincident with the \p decay time.
Timings in most of the veto counters were calculated by using the pulse shape near the peak;
with this method, possible errors in the timing evaluations due to overlapping multiple pulses,
which could give a detection inefficiency, was reduced.
The typical energy thresholds for the veto cuts were 2-3~MeV for the photon veto counters and 0.2~MeV for CV
to achieve tight rejection to both the extra photons and charged particles in the \KL decay.
The shape of the cluster in the $x$-$y$ plane was required
to be consistent with the expected shape of an electromagnetic shower due to a single photon obtained by simulation \cite{SatoPhD}.
This requirement rejected events containing a cluster made by multiple photons or a hadronic shower made by neutrons.
We also developed neural network cuts to further remove neutron contributions,
based on the difference of kinematic features and cluster shapes between photon and neutron in the calorimeter.
Details on the selection criteria are described in Ref.~\cite{MaedaPhD}.

\begin{figure}[tb]
	\begin{center}
		\includegraphics[width=0.7\textwidth,clip,trim=5 5 10 30]{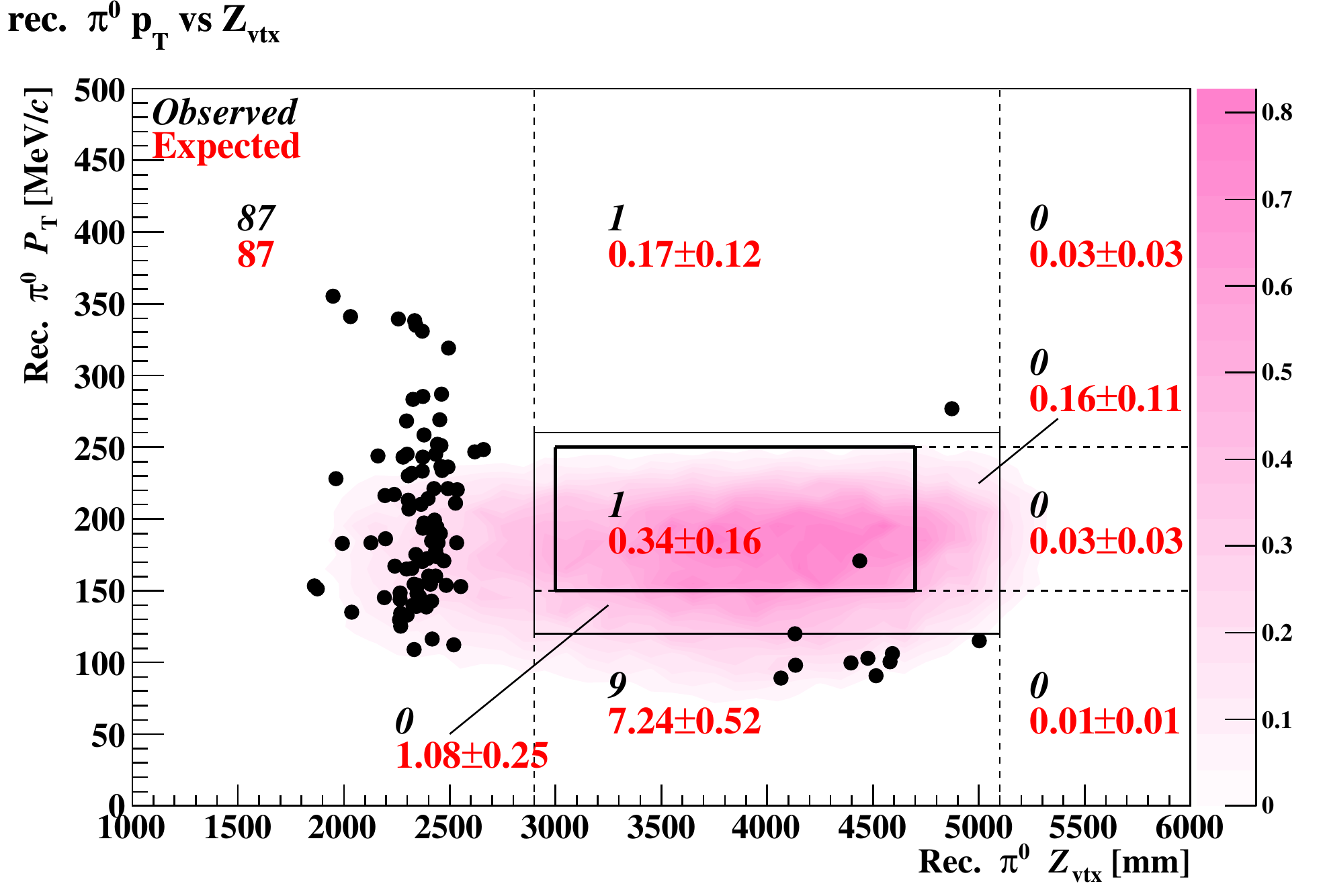}
         \end{center}
	\caption{Reconstructed \p transverse momentum (\pt) versus decay vertex position (\zvtx)
		of the events with all the analysis cuts imposed.
		The region surrounded with a thick solid line is the signal region.
		The black dots represent the data, and the contour indicates the distribution of the \kpnn decay from MC.
                 The events in the region surrounded with a thin solid line were not examined before the cuts were finalized.
                 The black italic (red regular) numbers indicate
                 the numbers of observed events (expected background events) for the regions divided by solid and dashed lines.}
	\label{fig:BG}
\end{figure}

We set the signal region for the \kpnn decay using the \pt and \zvtx of the reconstructed \p. 
The \pt was required to be larger than 150~MeV/$c$ to remove \kcppp events
for which the \p is restricted to have a transverse momentum less than 133~MeV/$c$ \cite{PDG}.
To avoid contaminations from halo neutron interactions with detectors,
we also required 3000~$<$~\zvtx~$<$~4700~mm.
The probability that a \KL entering the KOTO detector decays in this \zvtx region was 3.2\%.
The events observed in this analysis are shown in Fig.~\ref{fig:BG},
in which all the cuts except for \pt and \zvtx have been imposed.
To avoid introducing bias in the cut optimization,
all criteria were pre-determined before examining events in and around the signal region, indicated by the box with a thin solid line in Fig.~\ref{fig:BG}.

\section{Background estimation}
Table~\ref{tab:BG} summarizes the estimation of background events in the signal region.
The numbers of observed events and estimated background events
inside and outside of the signal region are also shown in Fig.~\ref{fig:BG}.
The backgrounds were categorized into two types: \KL backgrounds and neutron backgrounds. 

\begin{figure}[tb]
	\begin{minipage}{0.49\textwidth}
	(a)
	\begin{center}
		\includegraphics[width=\textwidth,clip,trim=25 30 55 35]{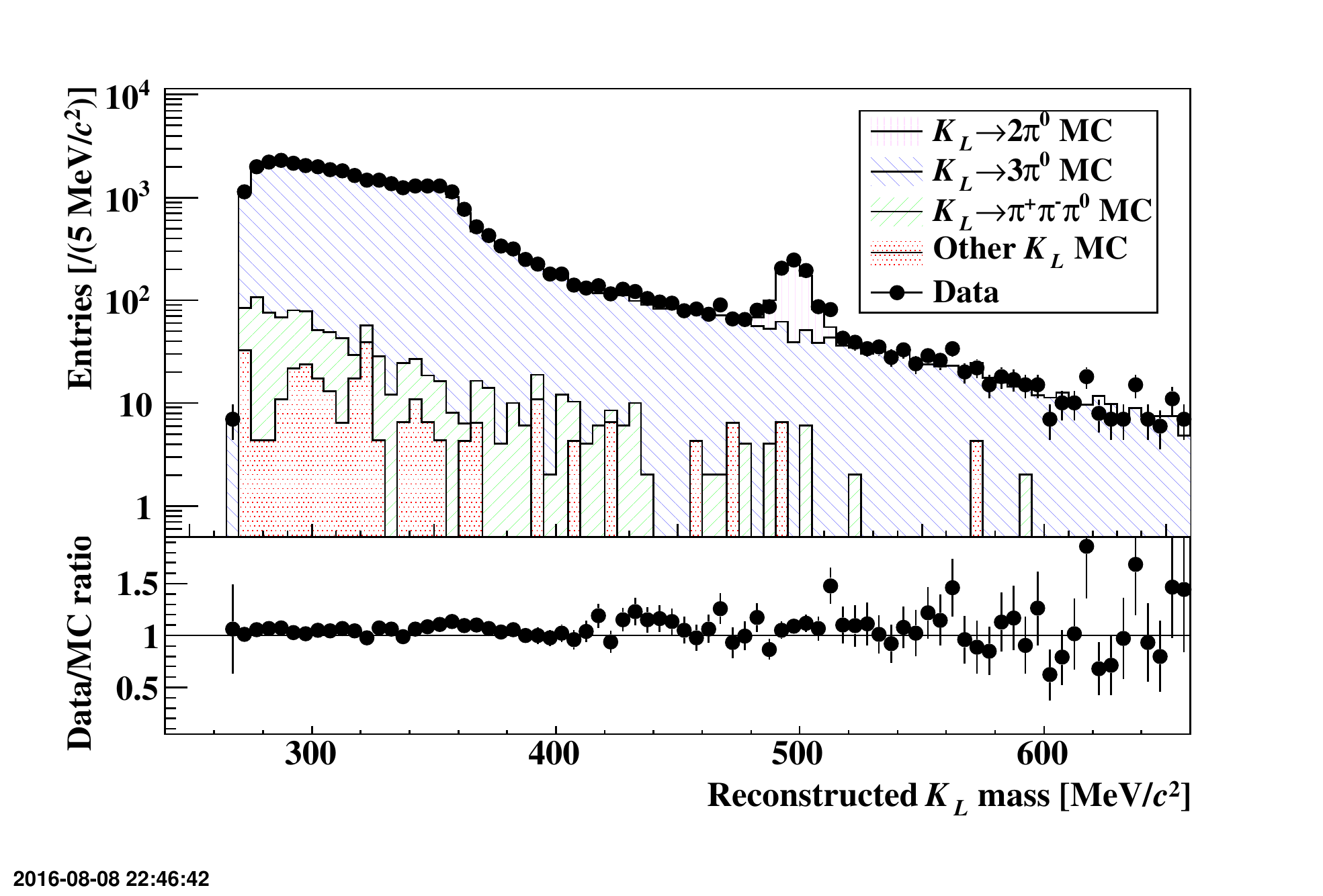}
         \end{center}
	\end{minipage}
	\begin{minipage}{0.02\textwidth} \: \end{minipage}
	\begin{minipage}{0.49\textwidth}
	(b)
	\begin{center}
		\includegraphics[width=\textwidth,clip,trim=25 30 55 35]{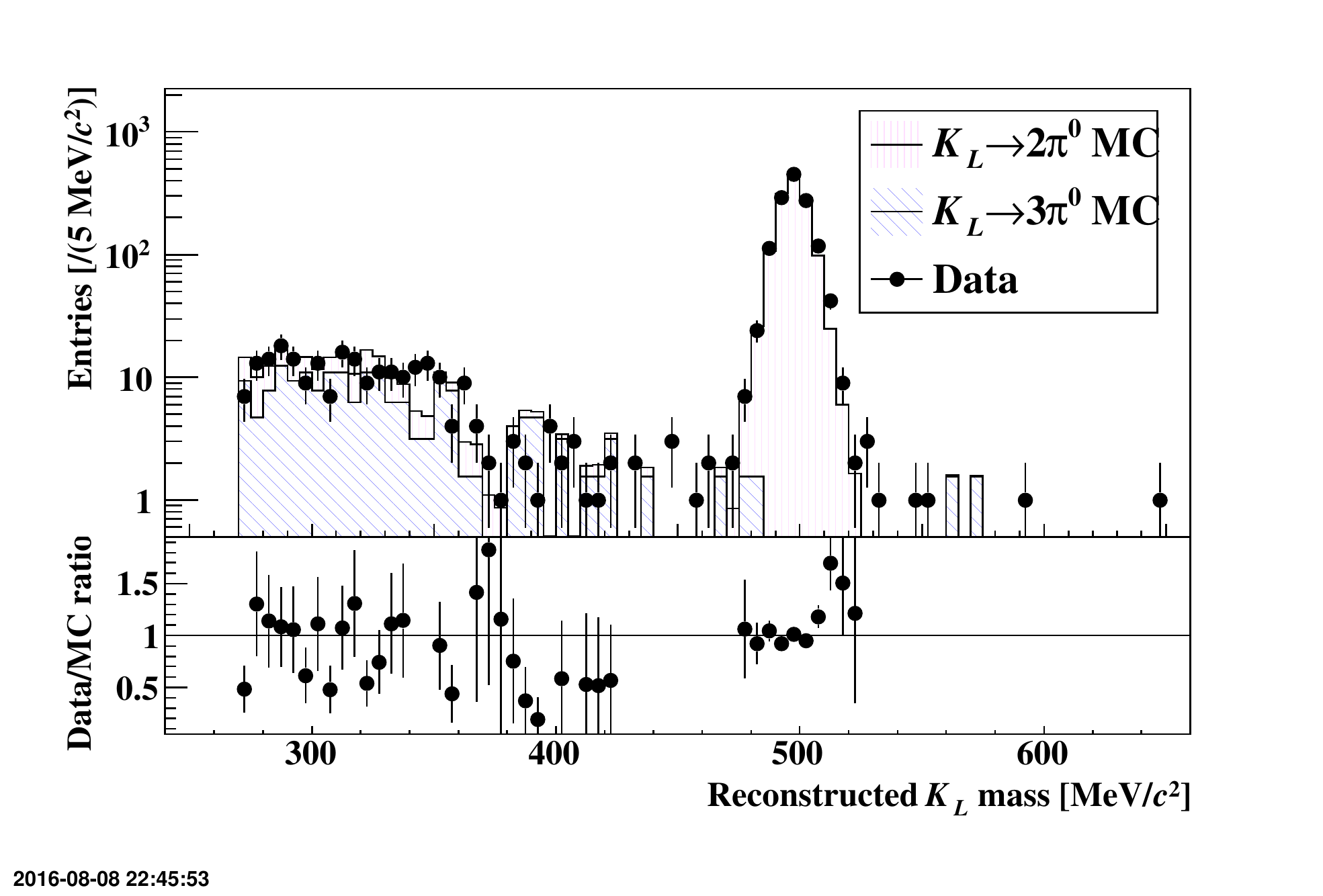}
	\end{center}
	\end{minipage}
	\caption{Distributions of four-photon invariant mass of the \kpp events before (a) and after (b) imposing the veto cuts.
		The dots with error bars are for data, and colored histograms are for MC.
		The bottom regions in both panels show the ratio of data and MC events for each histogram bin.}
	\label{fig:KLpi0pi0}
\end{figure}

\begin{figure}[tb]
	\begin{minipage}{0.49\textwidth}
	(a)
	\begin{center}
		\includegraphics[width=\textwidth,clip,trim=25 30 10 30]{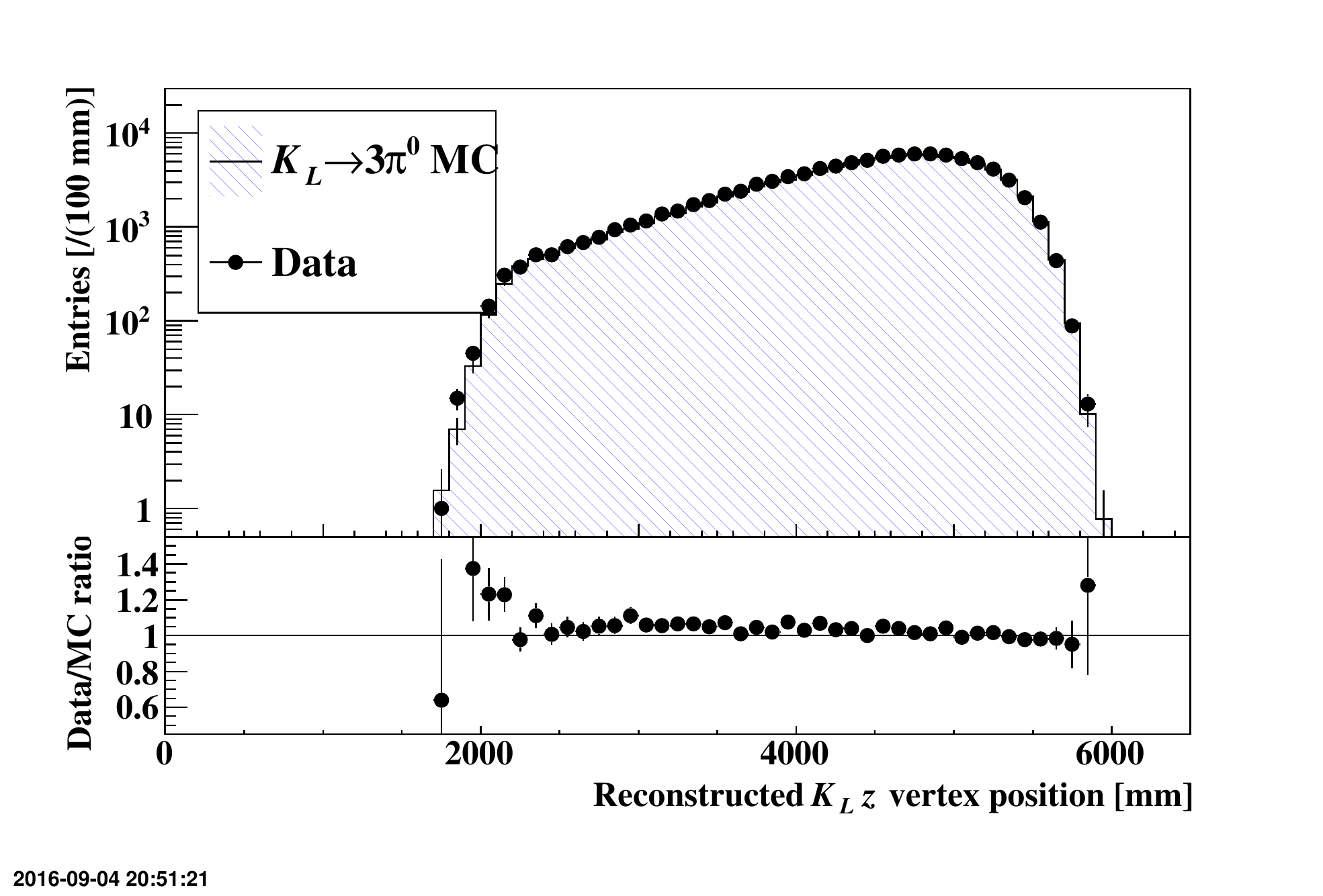}
        \end{center}
	\end{minipage}
	\begin{minipage}{0.02\textwidth} \: \end{minipage}
	\begin{minipage}{0.49\textwidth}
	(b)
	\begin{center}
		\includegraphics[width=\textwidth,clip,trim=25 30 10 30]{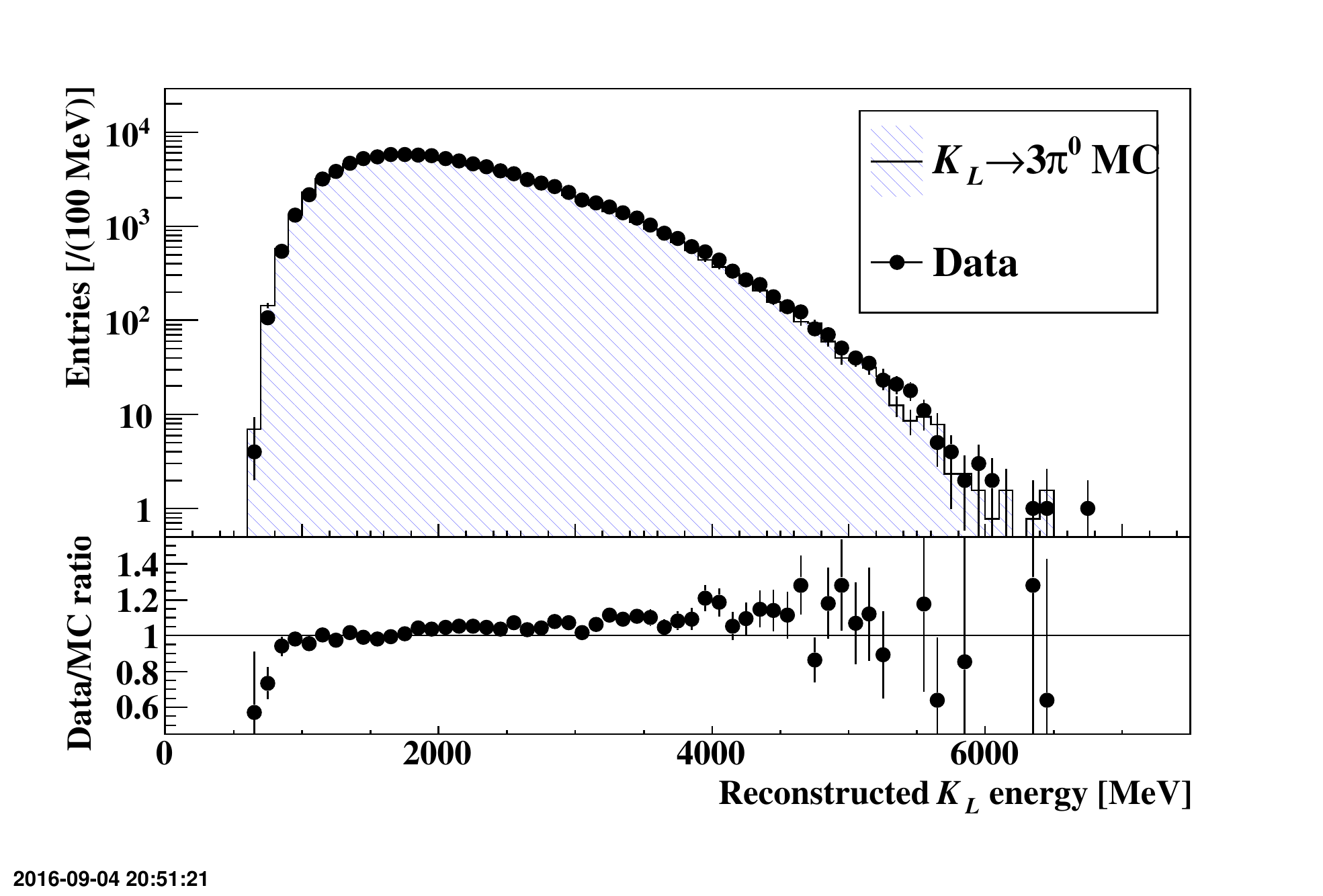}
	\end{center}
	\end{minipage}
	\caption{Distributions of reconstructed \KL decay vertex position (a) and energy (b) of the \kppp events.
		The bottom regions in both panels show the ratio of data and MC events for each histogram bin.}
	\label{fig:KL3pi0_Zvtx}
\end{figure}

The \KL backgrounds were evaluated by using Geant4-based \cite{Geant4} Monte Carlo simulations (MC) for each \KL decay mode.
Accidental hits in the KOTO detector had been collected simultaneously with the data taking, and were overlaid to the MC samples.
The background studies based on MC were validated with the \kpp and \kppp data.
To reconstruct the \kpp (\kppp) decays, events with four (six) photons in the calorimeter were selected.
Among all possible combinations of the photons,
the combination that had the best agreement of the \zvtx values for two (three) \ps was adopted.
Figure~\ref{fig:KLpi0pi0} shows the four-photon invariant mass distribution of the \kpp events before and after imposing the veto cuts.
The events in the low mass region are mostly from the \kppp decays in which four out of six photons were detected in the calorimeter.
The reduction of these events by detecting the extra two photons in the veto counters is well reproduced by the MC.
Figure~\ref{fig:KL3pi0_Zvtx} shows the distributions of the reconstructed \KL decay vertex position and energy of the \kppp events,
which indicated the acceptance derived from MC was well understood.

The \kpp decay is the major \KL background source because there are only two extra photons which can be detected by veto counters.
We generated a MC sample with 40-times the statistics of the data.
With two MC events that remained in the signal region after imposing all the cuts,
the background contribution was estimated to be 0.047~events.
For other decay modes,
we generated the MC samples with various assumptions of topologies or mechanisms that could cause backgrounds to \kpnn.
In the case of \kcppp, for instance, the decay can be a background
if charged pions hit the downstream beam pipe, which was made of 5-mm-thick stainless steel, and were undetected.
The nine events located in the low-\pt region (\pt~$<$~120~MeV/$c$, 2900~$<$~\zvtx~$<$~5100~mm) in Fig.~\ref{fig:BG}
are explained by this mechanism.
The requirement \pt~$>$~150~MeV/$c$ reduced the \kcppp background to a negligible level.
The \kgg decay can be a background if an incident \KL is scattered at the upstream vacuum window,
which was made of 125-$\mu$m-thick polyimide film, and obtains a finite transverse momentum.
The \kgg decay was simulated for \KLs with the transverse momentum larger than 20~MeV/$c$,
and the number of this background was estimated to be 0.03~events.
Accidental hits overlapping with the \ke, \kmu, \kppp, and \kcppp decays can change the reconstructed hit timing and cause an inefficiency.
The background due to this inefficiency was separately treated and estimated to be 0.014~events.

The neutron backgrounds were caused by halo neutrons. 
One type of neutron background came from halo neurons interacting with the NCC detector material in the upstream end of the decay volume.
Secondary particles by such interactions were detected by the calorimeter and mimicked the \p from the \kpnn decay.
The events in the upstream region (\zvtx$<$2900~mm) in Fig.~\ref{fig:BG} are \ps produced by this process.
They can be a background when photon energies are mismeasured or a secondary neutron is detected in the calorimeter
and misidentified as a photon.
The contribution due to this mechanism was estimated with the MC,
in which the halo neutrons were generated by a beam-line simulation and its yield was normalized to the number of the upstream events in the data.
We evaluated this background to be 0.056~events.

Another type of the neutron background was due to halo neutrons hitting directly the calorimeter.
A neutron incident on the calorimeter can deposit energy through hadronic interactions,
and a secondary neutron from these interactions can deposit energy at another place after traveling inside the calorimeter.
The contribution due to this mechanism was estimated with the data
from a special run with a 5-mm-thick aluminum plate inserted inside the beam at $z=2795$~mm.
Neutrons in the beam scattered at the plate would hit the calorimeter and mimic the background events.
We used the events which remained inside the signal region in this special run to estimate the number of background events.
This sample was also used for training the neural network used for making cuts in the analysis.
This background, estimated to be 0.18~events, was found to be the main background source in this analysis.

Details on the background estimation are described in Ref.~\cite{MaedaPhD}.
The total number of expected background events was $0.34\pm0.16$.

\begin{table}[t]
	\caption{Summary of background estimation in the signal region.}
	\label{tab:BG}
	\begin{center}
	\begin{tabular}{cc}
	\hline
	background source & number of events \\ \hline
	\kpp & $0.047\pm0.033$ \\
	\kcppp & $0.002\pm0.002$ \\
	\kgg & $0.030\pm0.018$ \\
	pileup of accidental hits & $0.014\pm0.014$ \\
	other \KL background & $0.010\pm0.005$ \\
	halo neutrons hitting NCC & $0.056\pm0.056$ \\
	halo neutrons hitting the calorimeter & $0.18\pm0.15$ \\ \hline 
	total & $0.34\pm0.16$ \\ \hline
	\end{tabular}
	\end{center}
\end{table}

\section{Normalization}	\label{sec_norm}
The number of observed events was normalized to the total number of the \KL decays in the collected data,
which was obtained from the \kpp decay events in the normalization data.
We calculated the single event sensitivity (SES), which corresponds to the branching fraction with which we expect one signal event, as
\begin{equation}
\notag
\mathrm{SES} = \frac{1}{A_{\mathrm{sig}}}\frac{A_{\mathrm{norm}}\;\mathrm{Br}(\kppEq)}{p\;N_{\mathrm{norm}}},
\end{equation}
where $A_{\mathrm{sig}}$ ($A_{\mathrm{norm}}$) is the acceptance for the \kpnn (\kpp) decay,
$\mathrm{Br}(\kppEq)$ is the branching fraction of the \kpp decay \cite{PDG},
$p$ is the prescale factor of 30 used to collect the \kpp events, and $N_{\mathrm{norm}}$ is the number of observed \kpp events in the data.
The acceptance was defined as the efficiency that two (four) photons from the \kpnn (\kpp) decay were detected by the calorimeter,
reconstructed as the \KL decay, and passed all the cuts.
Using the acceptance ratio between the \kpnn and \kpp decay modes reduced systematic uncertainties.
The acceptances for \kpnn and \kpp were evaluated with MC
as $A_{\mathrm{sig}} = 1.02\%$ and $A_{\mathrm{norm}} = 0.59\%$.
Based on the 1296 reconstructed \kpp events within $\pm$15~MeV/$c^{2}$ of the \KL mass peak shown in Fig.~\ref{fig:KLpi0pi0} (b),
the SES was obtained to be $(1.28\pm0.04_{\mathrm{stat.}}\pm0.13_{\mathrm{syst.}})\times10^{-8}$.
This value is comparable to the final sensitivity of the E391a experiment,
which was derived with a number of POT of $2.5\times10^{18}$ with 12-GeV protons over four months \cite{E391a}.

Systematic uncertainties in SES are summarized in Table~\ref{tab:NormSyst}.
The major uncertainties came from the veto cuts and the L2 trigger effect.
The uncertainty in the veto cuts was evaluated by comparing the extra-particle rejection factors,
defined as $N_{\mathrm{all}} / N_{\mathrm{cut}}$, of each veto counter between data and MC,
where $N_{\mathrm{all}}$ is the number of events after imposing all the cuts,
and $N_{\mathrm{cut}}$ is the number of events with all the cuts except a specific one.
The \kgg decay instead of \kpnn was used for the signal acceptance.
The BCV veto cut showed the largest uncertainty of 4.0\%.
The uncertainty in the L2 trigger effect was due to trigger bias that remained after the offline COE cut was imposed.
It did not affect the \kpp acceptance because the L2 trigger was not used for the normalization data.
The uncertainty was evaluated by studying the L2 efficiency as a function of the reconstructed COE value in the offline analysis
from data samples with various trigger requirements and cuts.
This uncertainty was found to be the largest, 6.56\%.

\begin{table}[t]
	\caption{Systematic uncertainties in the single event sensitivity.}
	\label{tab:NormSyst}
	\begin{center}
	\begin{tabular}{cc}
	\hline	
	source & relative uncertainty [\%] \\ \hline
	\KL momentum spectrum & $\pm1.51$ \\
	photon selection cut & $\pm1.07$ \\
	kinematic cuts for \kpp & $\pm2.46$ \\
	kinematic cuts for \kpnn & $\pm2.81$ \\
	cluster-shape-related cuts & $\pm2.51$ \\
	veto cuts & $\pm5.50$ \\ 
	L2 trigger effect & $\pm6.56$ \\
	MC statistics & $\pm0.78$ \\ 
	\kpp branching fraction \cite{PDG} & $\pm0.69$ \\ \hline
	total & $\pm$9.9 \\ \hline
	\end{tabular}
	\end{center}
\end{table}

\section{Results}
After all the selection criteria were determined,
the events in the signal region were examined.
One event was observed as shown in Fig.~\ref{fig:BG}.
The number is consistent with with the expected number of background events summarized in Table~\ref{tab:BG}.
Assuming Poisson statistics with the statistical and systematic uncertainties considered \cite{ULwithError},
the 90\% C.L. upper limit on Br$(\kpnnEq)$ was set to be $5.1\times10^{-8}$.

\begin{figure}[tb]
	\begin{center}
		\includegraphics[width=0.5\textwidth,clip,trim=10 5 55 36]{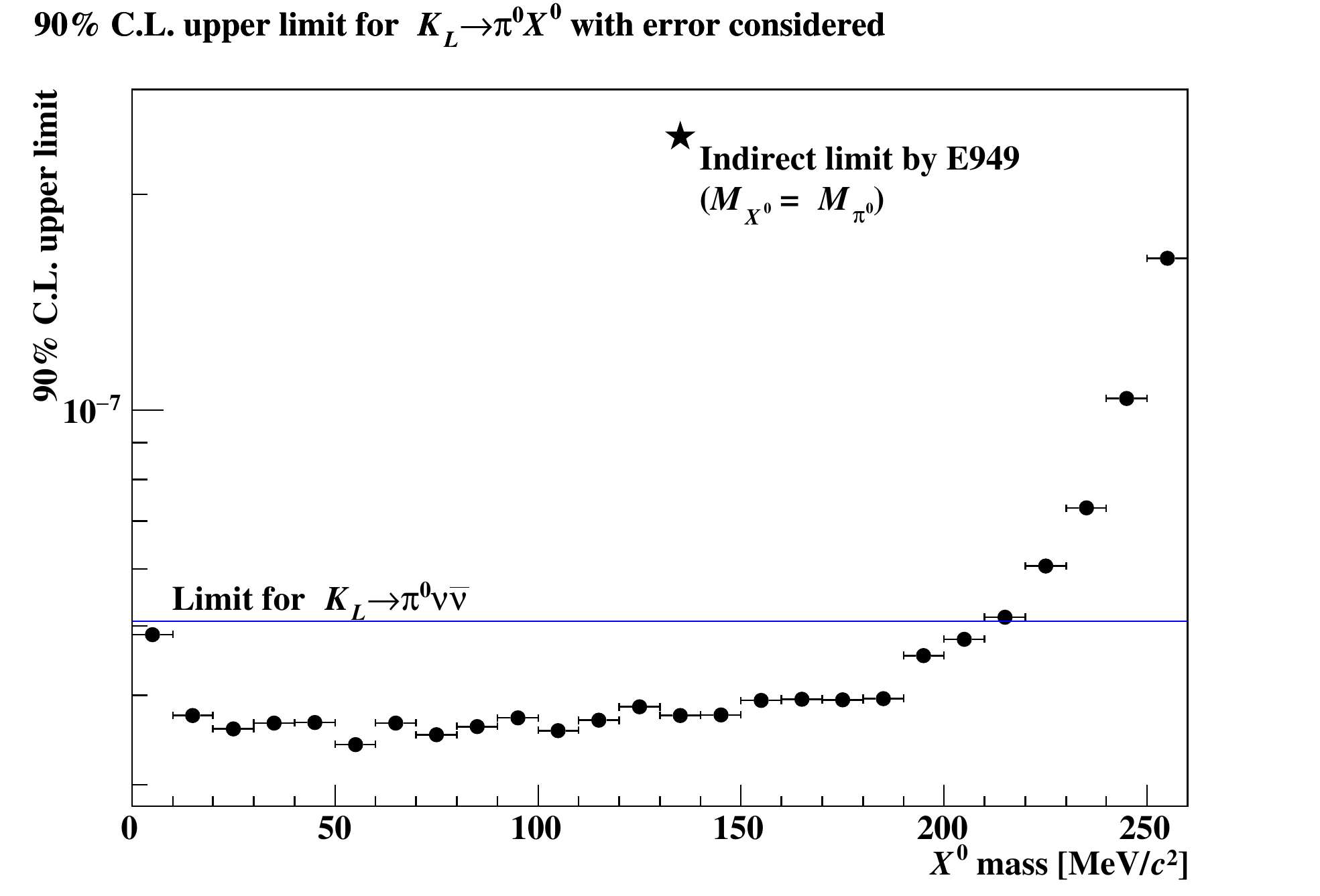}
	\end{center}
	\caption{Upper limit at the 90\% C.L. on the \kpx branching fraction as a function of the $X^{0}$ mass.
		For comparison, the limit on the \kpnn decay is shown with the blue line
		and the indirect limit by the E949 experiment with the star mark.}
	\label{fig:KLpi0X}
\end{figure}

We also studied the \kpx decay suggested in Ref.~\cite{KLpi0X, KLpi0X_new}.
The same selection criteria as the \kpnn search were used.
The acceptance was evaluated and the corresponding upper limit was obtained for each $X^{0}$ mass.
Figure~\ref{fig:KLpi0X} shows the 90\% C.L. upper limit for the \kpx decay as a function of the $X^{0}$ mass;
we set an upper limit $\mathrm{Br}(\kpxEq)<3.7\times10^{-8}$ at the 90\% C.L.
for the $X^{0}$ mass of 135~MeV/$c^{2}$.
This is the first result by a direct search for this process.
It also improves the indirect limit obtained by scaling the limit given by the BNL E949 experiment,
$\mathrm{Br}(K^{+}\rightarrow\pi^{+}X^{0})<5.6\times10^{-8}$ \cite{E949}, by a factor 6.5.

\section{Conclusions and prospects}
Based on the 100-hour data taken with a 24-kW beam power in the first physics run in 2013,
we set an upper limit on Br$(\kpnnEq)$ as $5.1\times10^{-8}$ at the 90\% C.L.
and on Br$(\kpxEq)$ as $3.7\times10^{-8}$ at the 90\% C.L. for the $X^{0}$ mass of 135~MeV/$c^{2}$.
This is the first direct search for the \kpx process.
Our result improves a previous indirect bound by a factor of 6.5 for $m_{X} = m_{\pi^{0}}$.

Since 2013, we have upgraded several detector subsystems to reduce \KL backgrounds \cite{BHGC, IB},
and in 2015, we collected 20 times more data than what is reported in this letter.
We also developed new analysis methods to discriminate photons from neutrons
based on shower cluster shape \cite{IwaiNIM} and waveform of hit crystals \cite{SugiyamaPhD}.
We are also preparing to modify the calorimeter to add a neutron discrimination capability \cite{NakagiriKaon}.
With these improvements, we expect to suppress the neutron background sufficiently for the future results.

\ack
We would like to express our gratitude to all members
of the J-PARC Accelerator and Hadron Experimental Facility groups for their support. 
We also thank the KEK Computing Research Center for KEKCC 
and National Institute of Information for SINET4. 
This material is based upon work supported by
the Ministry of Education, Culture, Sports, Science, and Technology (MEXT) of Japan
and the Japan Society for the Promotion of Science (JSPS)
under the MEXT KAKENHI Grant Number JP18071006,  
the JSPS KAKENHI Grant Number JP23224007 
and through the Japan-U.S. Cooperative Research Program in High Energy Physics;
the U.S. Department of Energy, Office of Science, 
Office of High Energy Physics, under Award Numbers
DE-SC0006497, DE-SC0007859, and DE-SC0009798; 
the Ministry of Education, the National Science Council/Ministry of Science
and Technology in Taiwan
under Grant Numbers 99-2112-M-002-014-MY3 and 102-2112-M-002-017;
and
the National Research Foundation of Korea 
(2012R-1A2A2A004554, 2013K1A3A7A06056592(Center of Korean J-PARC Users), and
 2015R1A2A1A15056254). 
Some of the authors were supported by Grants-in-Aid for JSPS Fellows.

\end{document}